\pdfoutput=1
\documentclass[journal]{IEEEtran}
\usepackage[utf8]{inputenc}
\usepackage[T1]{fontenc}
\usepackage[strings]{underscore}
\usepackage{graphicx}
\usepackage{booktabs}
\usepackage{amsmath,amssymb}
\usepackage{array}
\usepackage{tabularx}
\usepackage{multirow}
\usepackage{xcolor}
\usepackage{flushend}
\usepackage[font=footnotesize,labelfont=bf]{caption}
\usepackage{enumitem}
\setlist{nosep,leftmargin=1.3em}
\usepackage{listings}
\usepackage{textcomp}

\definecolor{codebg}{HTML}{F6F8FA}
\definecolor{navy}{HTML}{1E3A5F}
\definecolor{kw}{HTML}{1E40AF}
\definecolor{cm}{HTML}{6B7280}
\definecolor{st}{HTML}{B45309}

\lstdefinelanguage{SV}{
  morekeywords={module,endmodule,input,output,logic,wire,reg,always_ff,always_comb,
    posedge,negedge,if,else,begin,end,case,endcase,unique,typedef,enum,struct,packed,
    assign,localparam,parameter,for,int,bit,byte,signed,unsigned,function,endfunction,
    return,default,genvar,generate,endgenerate,always,initial,longint,shortint,break,
    automatic,import,package,endpackage,localparam},
  sensitive=true, morecomment=[l]{//}, morecomment=[s]{/*}{*/}, morestring=[b]"}
\usepackage{hyperref}
\hypersetup{colorlinks=true,linkcolor=navy,citecolor=navy,urlcolor=navy,
  pdftitle={The World Model Hardware Accelerator},pdfauthor={Shashank},
  pdfsubject={WMHA: a diffusion-transformer inference accelerator, technical report}}
\graphicspath{{figs/}}

\newcommand{\code}[1]{\texttt{\small #1}}

\begin{document}

\title{The World Model Hardware Accelerator}

\author{Shashank, AI Researcher%
\thanks{Manuscript prepared August 2026, revised September 2026. This is an independent,
self-funded research project. The architecture, register-transfer-level
(RTL) implementation, Universal Verification Methodology (UVM) environment
and functional results were developed on industry-standard simulators; the physical
measurements use the open-source SiliconCompiler/OpenROAD sky130 flow
proven on our earlier training and inference chips. Corresponding contact:
\texttt{sshashan@alumni.usc.edu}.}}

\markboth{Shashank: The World Model Hardware Accelerator (WMHA) Technical Report}%
{Shashank: The World Model Hardware Accelerator}

\maketitle

\begin{abstract}
Diffusion transformers invert the arithmetic that autoregressive decoding
made familiar. There is no token-by-token recurrence: every denoising step
is a full-sequence forward pass over static shapes, so the entire schedule
is known at compile time and the only serial dimension is the step count
itself. We exploit that structure in WMHA, a latency-first
diffusion-transformer inference accelerator: a very-long-instruction-word
sequencer issues four engines from one instruction word, a weight-stationary
$16{\times}16$ dual-dot array streams FP8 and BF16 contractions, and a
single-pass online-softmax attention pipeline keeps keys and values resident
through a skewed software pipeline. The design is specified in a frozen
micro-architecture document, implemented in synthesizable SystemVerilog,
and verified against a double-precision reference model by a UVM
environment whose acceptance criterion is semantic: the device must run a real denoising trajectory and reduce mean squared
error against a clean latent by at least a factor of ten. It does so by a
factor of $23$, at both synthesized configurations, with zero element
failures across $237$ million checked values. Eleven application
benchmarks built from published model shapes, including the original
diffusion-transformer configuration, run on the device and report measured
occupancy beside separately labelled projections. Five engines are taken to
routed layout in sky130 with parasitic-annotated timing and
measured-activity power; the full chip is synthesized, and the host limit
that stopped its place-and-route is quantified together with the machine
that would remove it.

\end{abstract}


\section{Introduction}

A diffusion transformer~\cite{peebles2023dit} does not decode. An autoregressive language model produces one token at a time. It therefore drags a sequential dependence through every layer of every step. A diffusion model~\cite{ho2020ddpm} evaluates the same network repeatedly over a latent of fixed shape. Each denoising
step is a complete forward pass; the steps are serial, but nothing inside a
step is. Sequence length, head count, head dimension and tile shapes are all
known before the first cycle executes.

That difference is not cosmetic, and it propagates all the way down to the
control path. If the shapes are static, the schedule is static, and a
compiler can place every memory movement and every contraction at a fixed
offset in an instruction stream. There is no need for the dynamic issue
machinery that dominates a general-purpose core, and no need to discover
dependences at run time that were fully determined at compile time. The
classical answer to that situation is a very-long-instruction-word (VLIW)
machine. The standard objection to VLIW concerns data-dependent control flow
that a compiler cannot see. Diffusion inference contains none. The objection
therefore lapses for this workload.

WMHA takes that position seriously. A single instruction word issues up to
four engines: a systolic array, a vector unit, and two independent
direct-memory-access engines. Dependences are expressed as an explicit wait
mask. No interlock discovers them at run time. Loop control is a zero-overhead
counter, so the back edge of an inner loop costs nothing. The measured control
cost is nine cycles per instruction word. On a representative program the
engines are busy for $99.4\%$ of wall time.

The workload also has an unusual numerical character that shapes the
verification problem. A diffusion sampler is iterative and contractive: it
starts from noise and converges toward a clean latent. Small numerical
perturbations at one step are not amplified without bound across steps, which
is what makes aggressive low-precision arithmetic viable at all. It also means that a purely structural correctness criterion can be satisfied by a device that is arithmetically defensible and semantically useless. Matching every tensor to a reference within a tolerance is such a criterion. We therefore froze a \emph{semantic} acceptance gate before any RTL existed. The device must execute a real sampling trajectory and demonstrably denoise. The criterion is a tenfold reduction in mean squared error against a known clean latent. That gate is the acceptance criterion of Section~\ref{sec:verif}.

\subsection*{Contributions}
\begin{itemize}
\item A latency-first diffusion-transformer inference accelerator. It is specified to a frozen micro-architecture document and implemented in synthesizable SystemVerilog. It is verified by a UVM environment against a double-precision reference, and taken through an open-source physical flow to routed layout with parasitic-annotated timing and measured-activity power.
\item A semantic acceptance criterion for a generative accelerator, frozen
before implementation, and the argument for why element-wise tolerance
checking cannot replace it.
\item A software-pipelining result on the canonical program. Restructuring the instruction schedule alone, with no RTL change, raises multi-engine concurrency from $1.83\%$ to $44.66\%$ of cycles. The measured speedup is $1.484\times$, with the arithmetic proven unchanged.
\item Eleven application benchmarks built from published model shapes, video, robotics, protein, 3D asset, guidance and image, each running a real program on the device with hardware counters read back. Two of them falsified claims in our own workload analysis: batching a guidance pair does not amortise the weight stream on this program, and the original diffusion-transformer head dimension does not fit the fused attention pipeline at this on-chip capacity.
\end{itemize}

\section{Background and Workload Analysis}
\label{sec:background}

\subsection{Why diffusion inference is not decoding}
A denoising step evaluates a transformer over a latent of $N$ tokens with
model width $d$, $h$ heads and head dimension $d_h$. For the image-class
configuration we target~\cite{peebles2023dit,esser2024sd3}, $N{=}128$, $d{=}256$, $h{=}4$, $d_h{=}64$; a
video-class configuration~\cite{wan2025,kong2024hunyuanvideo} extends $N$ and the head count. The same block also serves diffusion policies for robot control~\cite{chi2023diffusionpolicy}, the diffusion head of AlphaFold~3~\cite{abramson2024af3} and vector-set 3D asset generation~\cite{zhang2023vecset}, which is why the application benchmarks of Section~\ref{sec:appbench} span those classes. Every step reads
the same weights. With $T$ steps and batch $B$, each weight byte is reused $T\times B$ times more than in a single forward pass. A weight-stationary dataflow~\cite{jouppi2017tpu} is therefore the obvious choice. The cost of loading a stationary operand amortizes over an entire trajectory.

The serial dimension is $T$. Per-step latency is therefore the product
metric, not aggregate throughput: halving per-step latency halves
time-to-image, whereas doubling batch throughput does not. This is the central reason WMHA is organized around keeping one step's critical path short. It is also why the software-pipelining result of Section~\ref{sec:perf} matters more than an equivalent throughput gain would.

A second consequence is that the latent stays resident. Across the whole trajectory the working state is one latent tensor plus whatever the current layer needs. There is no growing key/value cache as in autoregressive decoding. On-chip capacity is therefore sized by the layer working set.

\subsection{What the workload removes and adds}
Relative to a training accelerator, diffusion inference removes the
optimizer, all backward dataflows, and gradient storage. We also removed
on-chip random number generation: stochastic samplers consume noise tiles,
and a host can supply them as ordinary tensors, so the entropy source need
not be silicon. This is a deliberate leanness argument, hardware that exists
must be verified, and a noise generator would have added a verification
surface for no architectural benefit.

What the workload adds is a different operator set. It needs SiLU and GELU piecewise-linear units, adaptive layer-norm modulation, and classifier-free guidance combination~\cite{ho2022cfg}. It also needs sampler updates of the form $z \leftarrow az+b$ with host-supplied scalars, which cover the DDPM and DDIM families~\cite{ho2020ddpm,song2021ddim} and the learned-covariance variant~\cite{nichol2021improved}. Arithmetic is FP8 E4M3 and BF16 with FP32 accumulation~\cite{micikevicius2022fp8}. These are individually cheap and collectively
decisive, an accelerator that streams contractions beautifully but returns
to a host for modulation would spend its time on the interconnect.

\subsection{The attention decision}
The single most consequential architectural choice was attention strategy.
A two-pass formulation computes all scores, normalizes, then multiplies by
values, and must store an $N\times N$ score matrix. A single-pass
online-softmax formulation~\cite{milakov2018online,rabe2021selfattention,dao2022flashattention} carries a running maximum and denominator,
rescaling partial outputs as it proceeds, and needs only a group of scores
resident at a time.

We chose the single-pass form for three reasons. It removes the quadratic
intermediate entirely. It lets keys and values stay resident in the
stationary region across a query stripe, which matches the weight-stationary
dataflow the array already implements. It converts attention into the same
streaming pattern as a feed-forward contraction, so one set of mechanisms
serves both.

The cost is a rescale chain. Every time the running maximum increases, the
accumulated output and denominator must be rescaled by a factor
$\alpha = 2^{m_{\text{old}} - m_{\text{new}}}$. That chain is numerically delicate in exactly the regime this workload produces: saturated low-precision scores drive the softmax toward one-hot.

\section{Architecture}
\label{sec:arch}

\subsection{Instruction model}
Source identifiers in this paper carry the \code{dix\_} prefix used in the
register-transfer-level tree; they name the same modules described here.

\begin{figure*}[t]\centering
\includegraphics[width=0.92\textwidth]{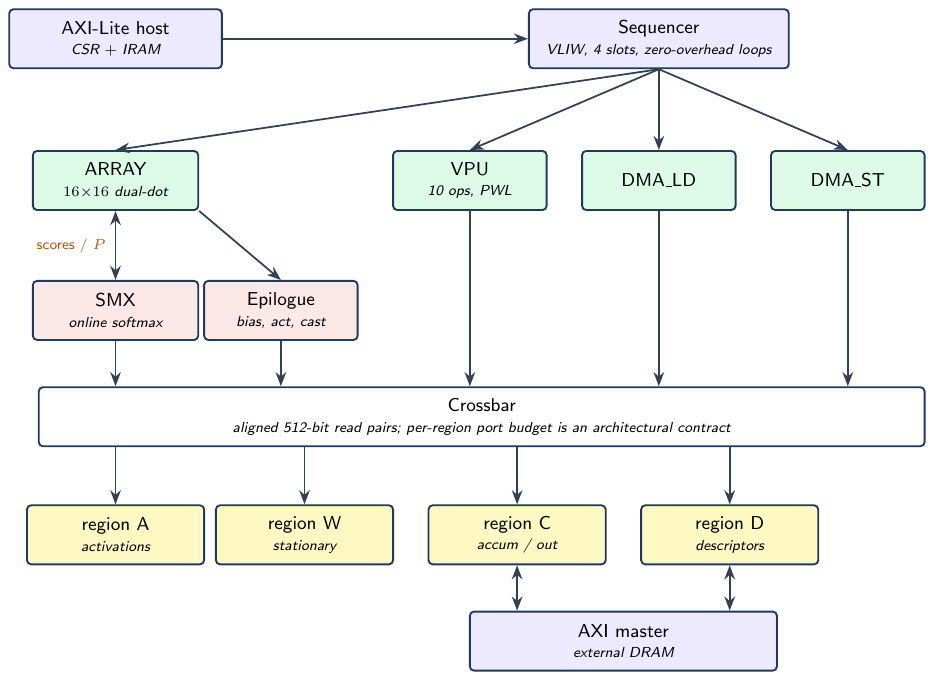}
\caption{WMHA top level. One VLIW sequencer issues four engines; the softmax
and epilogue units sit on the array's output path; all traffic reaches four banked regions through a crossbar whose per-region port budget is an architectural contract.}
\label{fig:top}\end{figure*}

One WMHA instruction word~\cite{fisher1983vliw} issues up to four engine slots: ARRAY (systolic
contraction), VPU (vector and activation), DMA\_LD and DMA\_ST. Each slot
carries an opcode, a reference to a tile descriptor, an immediate field, and
a \emph{wait mask} naming the engine classes whose in-flight work must retire
before this slot may issue. Dependences are thus explicit and resolved at
compile time; there is no scoreboard, no renaming, and no dynamic
disambiguation. The instruction memory holds $256$ words. That is a hard architectural budget: a program that does not fit does not run. The canonical whole-network program was sized against this limit from the first architectural estimate.

The array slot supports four operations. These are a forward contraction
(\code{G\_FWD}), two attention contractions (\code{A\_QKT} for scores and
\code{A\_PV} for the value product), and an output drain (\code{A\_DRAIN}).
The vector slot supports ten. They span layer and RMS normalization, adaptive
modulation, residual addition, the SiLU and GELU activations, rotary position
embedding, classifier-free guidance combination, a scaled-add sampler update,
and format casting.

Two properties of this encoding matter for everything that follows. First,
descriptor capture happens \emph{at issue}: an engine latches its descriptor
when the word issues, so a later word may overwrite that descriptor entry
without creating a hazard. This is precisely what permits a software pipeline
to stage iteration $i{+}1$'s descriptor while iteration $i$ is still draining.
Second, loop control is a zero-overhead construct. The back edge of an inner loop is taken inside the advance stage and costs no cycles. A tightly rolled program therefore pays no branch penalty, and a compiler has no reason to unroll for control alone.

\begin{figure}[t]\centering
\includegraphics[width=\columnwidth]{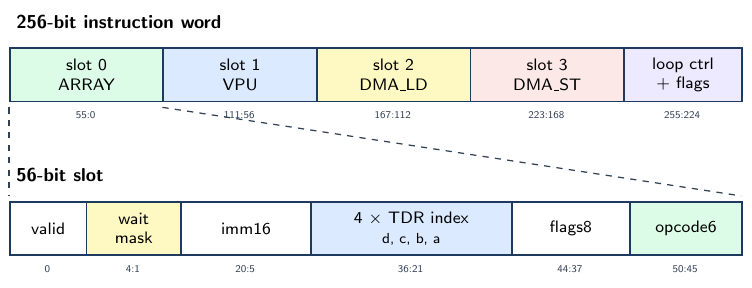}
\caption{The 256-bit instruction word: four 56-bit engine slots plus a
loop-control tail, with the slot expansion below. The \code{wait\_mask} names
the engine classes whose in-flight work must retire before this slot issues,
so dependences are explicit and resolved at compile time. Descriptor capture
happens \emph{at issue}, which is what lets a software pipeline stage
iteration $i{+}1$ while iteration $i$ is still draining.}
\label{fig:word}\end{figure}

\subsection{Contraction array}
\begin{figure}[b]\centering
\includegraphics[width=0.9\columnwidth]{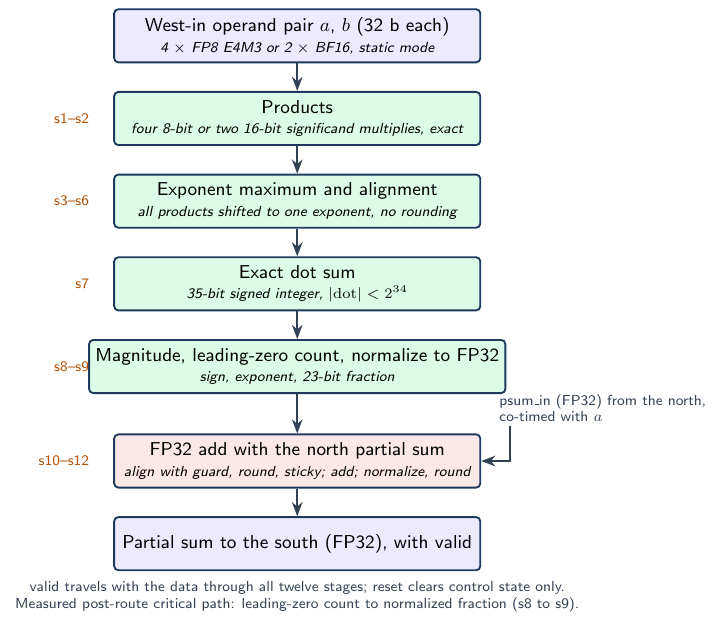}
\caption{One processing element (\code{dix\_pe\_cell.sv}), twelve register
stages. Four FP8 or two BF16 products are summed exactly as a 35-bit integer
before a single normalization to FP32, so the dot itself introduces no
rounding; the only rounding is the FP32 accumulation with the partial sum
from the north. The measured post-route critical path is the leading-zero
count feeding the normalized fraction.}
\label{fig:pe}\end{figure}

The array is a $16{\times}16$ grid of dual-dot cells (Fig.~\ref{fig:pe}), a weight-stationary systolic organization in the tradition of~\cite{jouppi2017tpu}. Each cell computes
either four FP8 (E4M3) products or two BF16 products as an exact dot,
accumulating into FP32. The cell was selected by an architecture-exploration bake-off across FP8, BF16 and INT8 candidates. Each candidate was evaluated by a real synthesis run. Analytical area models were avoided because the relative cost of a multiplier array and its accumulation tree is exactly the quantity they estimate worst.

A physical-design configuration halves the array to $8{\times}8$. Both
configurations are independently verified and every functional and coverage
result in this paper is reported for both. This is deliberate. A configuration that will be synthesized must pass on its own terms.

Operations may span multiple passes when the stationary operand exceeds one
array image, in which case the pass count is derived from the descriptor and
the accumulation is finalized automatically. Retirement is defined by an explicit table. An operation is in flight from issue until its final output write has landed. A dedicated deposit interlock prevents a drain from observing a partially written accumulator.

\subsection{Memory system}
On-chip memory is partitioned into four regions, activations, stationary
weights, accumulator/output, and descriptor space, each banked and reached
through a crossbar granting aligned $512$-bit read pairs. Each region has a fixed port budget, typically two reads and one write per cycle. That budget is an architectural contract. A program that requests more is committing a program error, and the hardware raises it as such.

Descriptors are walked by hardware. A descriptor carries a base address, a
stride, and a window; an advance steps the base by the stride and wraps
within the window. This gives a program a ring buffer without instruction
overhead, and it is what allows double-buffering of stationary operands
underneath a running contraction. Walker commitment follows the same
capture-at-issue rule as the rest of the machine.

\subsection{Attention pipeline}
\begin{figure*}[t]\centering
\includegraphics[width=0.78\textwidth]{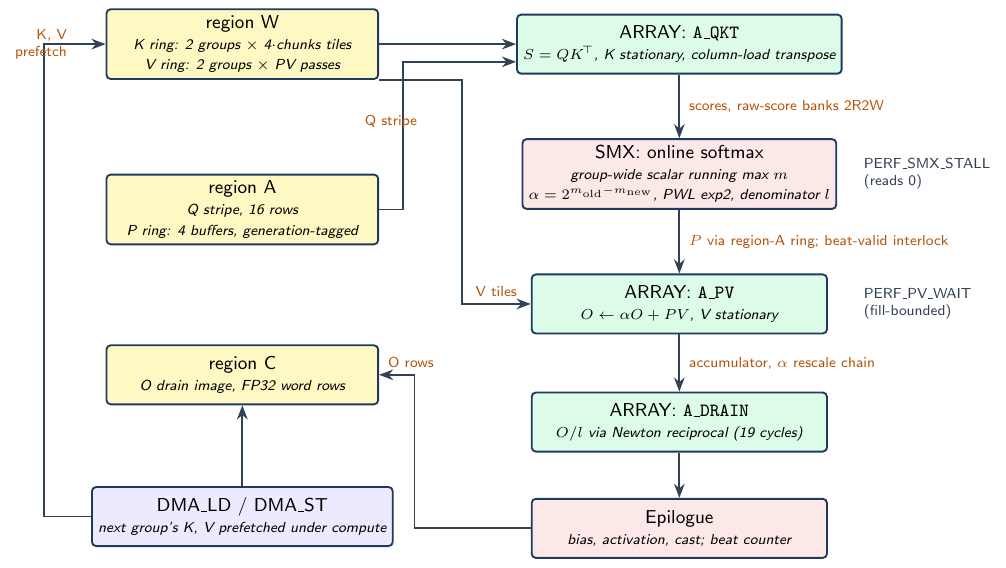}
\caption{Attention datapath for one key/value group. Keys and values sit in
two resident rings in region W; the query stripe and the generation-tagged
probability ring sit in region A. Scores leave the array for the online
softmax unit and return as probabilities under a beat-valid interlock; the
drain divides by the running denominator through a Newton reciprocal and
hands rows to the epilogue. Two counters make the interlock observable.}
\label{fig:attnpath}\end{figure*}

\begin{figure}[t]\centering
\includegraphics[width=\columnwidth]{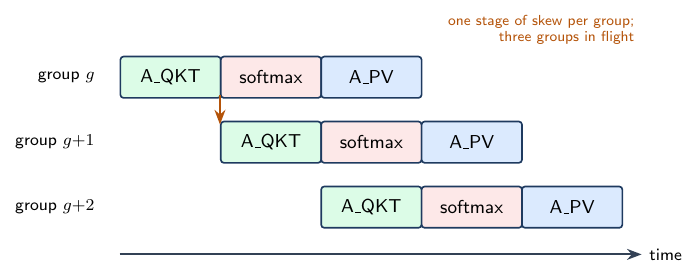}
\caption{Software-pipelined attention. Groups are skewed by three stages so a
score contraction, a softmax, and a value accumulation are in flight
simultaneously.}
\label{fig:attn}\end{figure}

Attention runs as a software-pipelined loop over key/value groups (Fig.~\ref{fig:attnpath}). Scores for a group are contracted on the array. An online-softmax unit maintains a running maximum and a running denominator and exponentiates through a piecewise-linear unit. The resulting probabilities stream back into the array to accumulate against values. Keys and values for the resident groups
stay in the stationary region across an entire query stripe, so the quadratic
score matrix never exists.

Three details are load-bearing. The running maximum is \emph{group-wide and scalar}, so one rescale factor serves the whole group. A beat-wise valid interlock couples the probability producer to the value consumer, and a dedicated counter measures consumer wait, so pipeline health is observable. The probability ring is generation-tagged, so a consumer can never read a previous group's probabilities from an aliased slot.

\section{Micro-architecture and Specification Discipline}
\label{sec:uarch}

\subsection{Freezing before building}
The micro-architecture specification was written and frozen before RTL
existed, and subjected to five rounds of adversarial review. Each round had three parts. A binding architect brief resolved the previous round's findings. A targeted rewrite applied them. A check then regressed the prior findings \emph{and} swept the newly introduced mechanisms with fresh eyes.
The finding count fell as shown in Table~\ref{tab:rounds}.

\begin{table}[t]
\centering\scriptsize
\setlength{\tabcolsep}{3pt}
\caption{Five-round adversarial specification review.}
\label{tab:rounds}
\begin{tabular}{lrl}
\toprule
round & findings & representative blocking finding \\
\midrule
1 & 78 & operand bandwidth undercounted $2\times$ in crossbar tables \\
2 & 29 & walker cannot ping-pong or wrap as specified \\
3 & 21 & commit-at-retire walker rule deadlocks the pipeline \\
4 & 14 & score-to-value pipeline claim false; $220$-cycle bubble \\
5 & 8 $\rightarrow$ 2 & drain latency missing a ceiling at ragged head dim \\
\bottomrule
\end{tabular}
\end{table}

The highest-yield review activities were not general reading. In descending order of blocking findings produced, they were as follows. Hand-assembling a program against the encoding tables. Computing per-bank port-grant arithmetic for the worst-case cycle. Walking the latency of the deepest pipelined schedule. Checking walker reachability for every descriptor-advance mode. Prose review
of the same document produced almost nothing by comparison.

\subsection{Pre-committed fallbacks}
One device is worth naming separately. Where a mechanism was uncertain, the
review pre-committed a fallback \emph{before} performing the analysis. The
attention schedule was proposed with a two-stage skew and a three-stage
fallback approved in advance. The two-stage form then failed its honest
derivation, the bubble it implied could not be hidden, and the fallback
fired without renegotiation.

We consider this more than bookkeeping. An architect who has already agreed what happens on failure has no incentive to rescue an attractive mechanism with optimistic arithmetic. That incentive is the failure mode that produces specifications which cannot be implemented as written.

\subsection{Errata as the only post-freeze edit}
Freezing changed the character of every later disagreement. After the freeze,
a mismatch between RTL and specification is not a discussion: either the RTL
is wrong, or the specification takes a dated erratum with a justification.
Twenty-two errata were issued. Their distribution is itself informative. Most correct a specification that was merely under-specified. One narrowed an alignment requirement that was correct at one configuration and wrong at the other.

\subsection{Pipeline structure and pinned latencies}
\begin{figure}[t]\centering
\includegraphics[width=0.96\columnwidth]{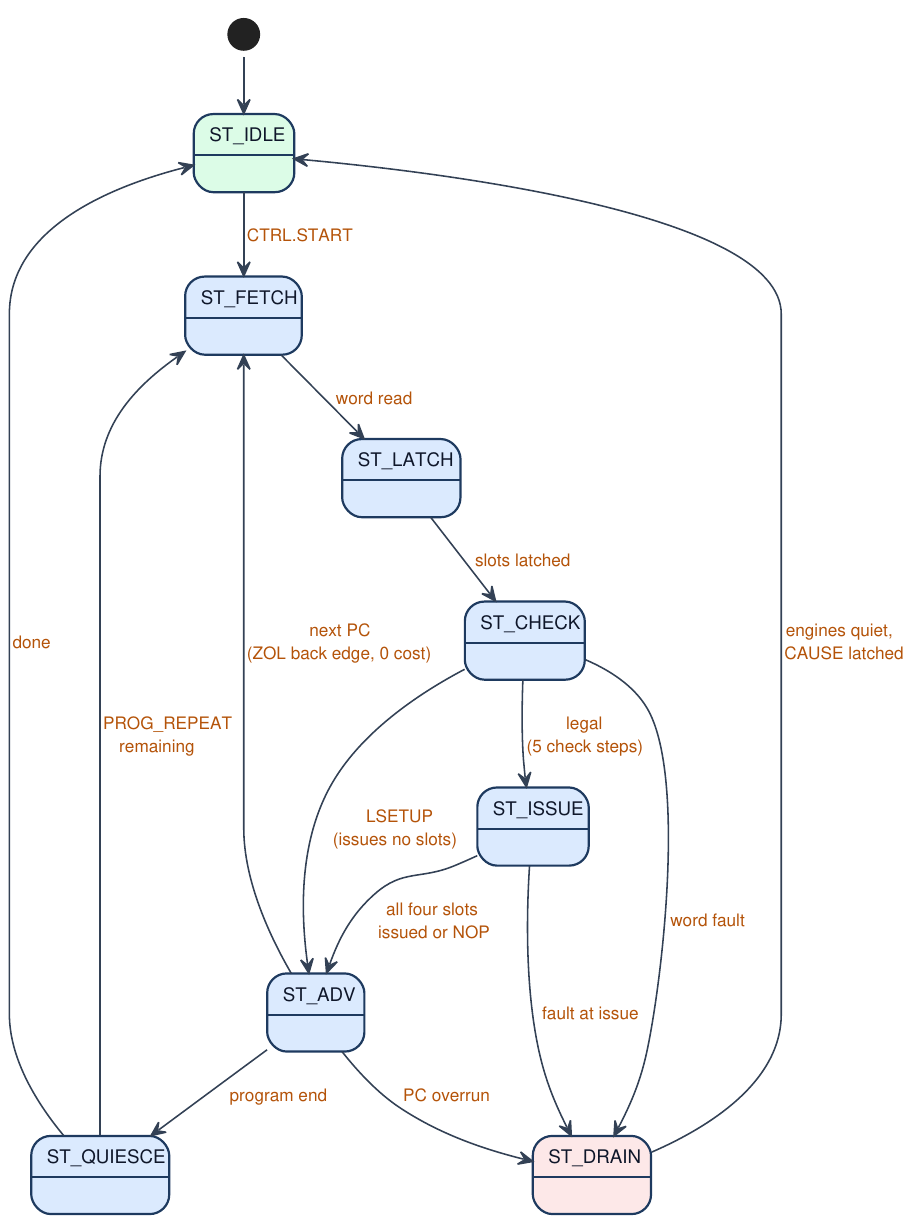}
\caption{Sequencer state machine (from \code{dix\_seq.sv}). Word atomicity is
enforced in \code{ST\_CHECK}, which faults the whole word before any slot
issues; capture-and-advance of descriptors happens at \code{ST\_ISSUE}. The
\code{ST\_ADV} back edge is the zero-overhead loop and costs no cycles.}
\label{fig:seqfsm}\end{figure}

\begin{figure}[t]\centering
\includegraphics[width=0.96\columnwidth]{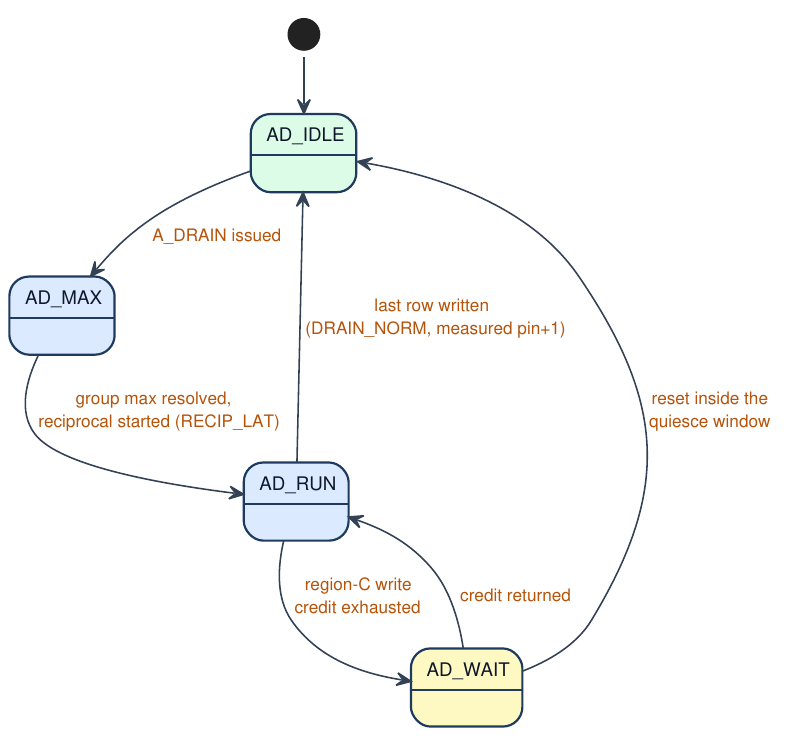}
\caption{Attention output-drain state machine. Its exit latency is the pinned
constant \code{DRAIN\_NORM}, measured uniformly over $51$ attempts at both
configurations.}
\label{fig:drainfsm}\end{figure}

\begin{figure}[t]\centering
\includegraphics[width=0.96\columnwidth]{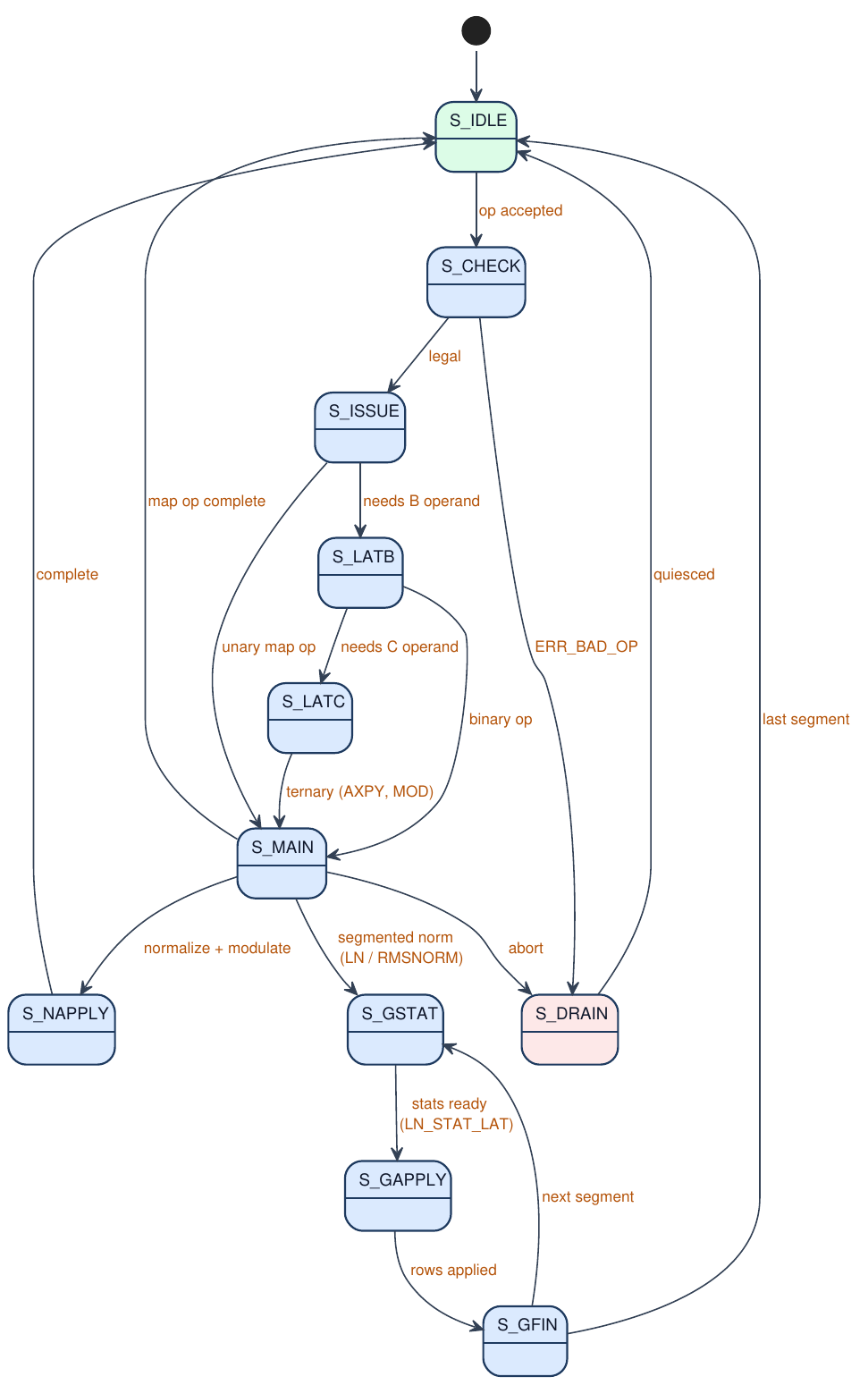}
\caption{Vector-unit state machine (from \code{dix\_vpu.sv}). Operand
latching is staged by arity (\code{S\_LATB}, \code{S\_LATC}), map operations
complete from \code{S\_MAIN}, and the segmented normalizations loop through
statistics, apply and finish states once per segment, with the statistics
tail pinned as \code{LN\_STAT\_LAT}. An illegal opcode faults the operation
before any operand is read.}
\label{fig:vpufsm}\end{figure}

\begin{figure}[t]\centering
\includegraphics[width=0.96\columnwidth]{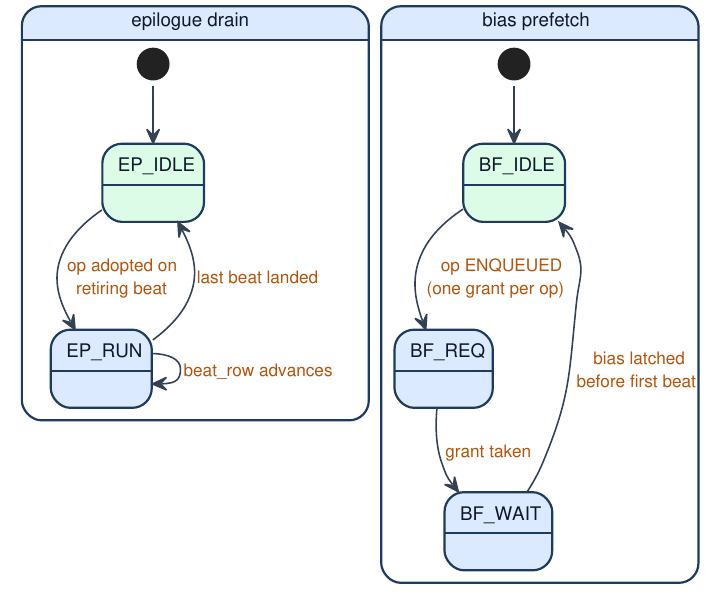}
\caption{Epilogue drain and bias-prefetch state machines. The drain stream
has no backpressure, so bias is fetched when an operation is enqueued and a
beat can never arrive before its bias is resident; the row counter advances
only on accepted beats.}
\label{fig:epifsm}\end{figure}

The engines expose their latencies as pinned architectural constants (Figs.~\ref{fig:seqfsm}--\ref{fig:epifsm}) so a
compiler can schedule against them. Representative values at the $16{\times}16$ configuration follow. An array operation is in flight for $14N{-}1 = 223$ cycles from injection through drain. The epilogue adds $9$. The reciprocal used by attention normalization takes $19$. The drain normalization path was measured at $150$ cycles for the widest head dimension.

These are contracts, not documentation. Where a constant is pinned, an assertion binds the RTL to it. Every such assertion now also reports its attempt count into the run summary.

\section{Numerics and the Tolerance Model}
\label{sec:numerics}

WMHA computes in FP8 (E4M3) and BF16 with FP32 accumulation, and is verified
against a double-precision reference. Almost no result is bit-exact, so almost every check is a tolerance check. The tolerance model is therefore part of the design.

\subsection{Composing a band}
Each recorded value carries metadata describing how it was produced. The metadata holds four things. The depth of the reduction tree the value emerged from. The length of any accumulation chain. A relative envelope for each piecewise-linear unit it passed through. An absolute term for storage quantization. A comparison composes these
into a band around the reference value. The composition is conservative by construction: an $\ell_1$ walk over the operations actually performed. It is therefore loose where cancellation would in practice reduce error. We accept that asymmetry deliberately. A bound that can be wrong in the permissive direction is not a bound.

Two carve-outs keep the model honest. First, some results are declared
\emph{exact-class} and compared bit-for-bit. A pure data movement must not be
granted a tolerance; if a DMA copy is allowed to differ from its source, the
check has stopped testing the copy. Second, values that the hardware's FP32 units flush to zero carry an explicit floor. The double-precision reference does not flush, and would otherwise demand a precision the format cannot represent.

\subsection{Storage saturation and its ceiling}
FP8 E4M3 saturates at $\pm448$. A value whose ideal exceeds that is stored
clamped, so the band carries the saturation ceiling as well as the
flush-to-zero floor. The ceiling term is identically zero inside the
representable range and prices the clamp outside it, so a correctly clamped
device output is never reported as an error.

\subsection{Propagation without double counting}
The remaining subtlety is composition across programs. When one program's
output feeds another, its band must propagate, or the consumer is checked
against a precision its input never had. A band that propagates must not also be counted a second time at the destination. A chain of $k$ tensors would otherwise accumulate its own error $k$ times over, amplified through every contraction between them.

We therefore separate each band into a self-injected part and a propagated
part. The element-wise comparison uses the full band; any composition across
a chain of re-baselined programs uses only the self-injected portion. The normalizations propagate through their Jacobian, which is bounded by the
reciprocal standard deviation of the row times $(1 + 2/d)$. For a row with a
small norm that factor is in the thousands, so a band that omitted it would
reject correct outputs at exactly such rows.

\subsection{The negative control}
A tolerance model that is too loose is worse than no model, because it
produces green results. We therefore maintain a negative control: a deliberately perturbed replay that the band must reject. Its margin is \emph{measured}. It currently misses its budget by a
factor of $2131$ against a required floor of $100$.

We regard a live, measured negative control as the minimum evidence that a
tolerance model is still doing work.

\section{Verification}
\label{sec:verif}

\subsection{Environment}
\begin{figure}[t]\centering
\includegraphics[width=0.92\columnwidth]{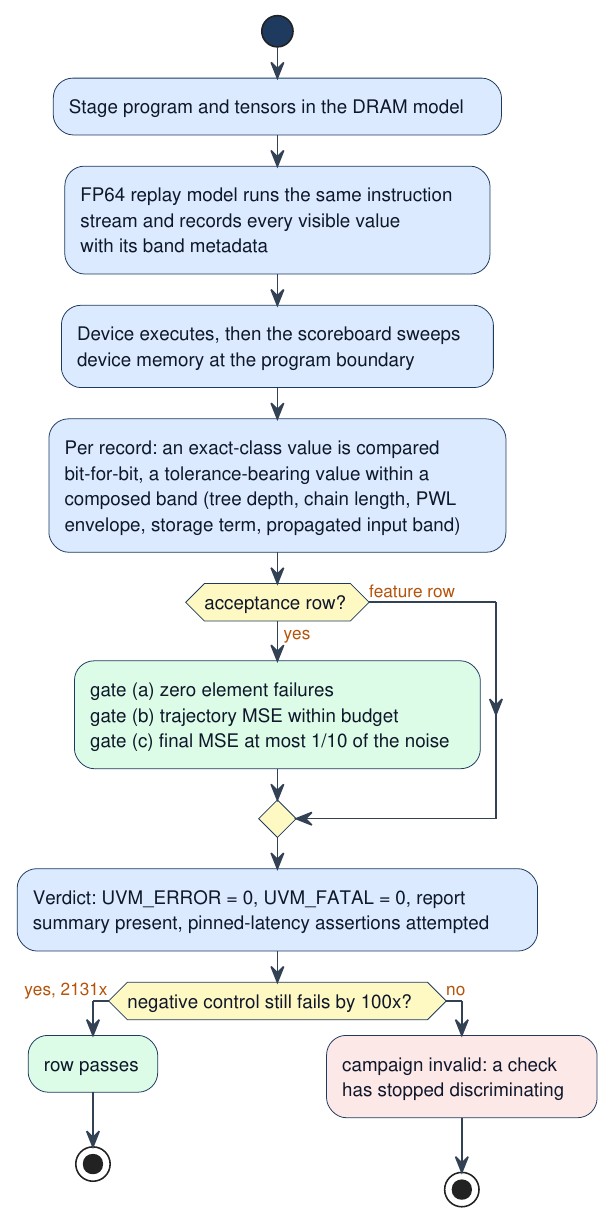}
\caption{Flow of one test row. The replay model records values with band
metadata, the scoreboard sweeps at program boundaries, the acceptance row
adds the three frozen gates, and a verdict step refuses a run whose log
lacks a report summary or whose pinned-latency assertions never attempted.
The negative control must keep failing by a measured margin, or the campaign
is declared invalid.}
\label{fig:verifflow}\end{figure}

A UVM~\cite{ieee18002uvm} environment (Fig.~\ref{fig:verifflow}) drives WMHA against a \emph{replay} reference model. The model executes the same instruction stream the device does, in double precision. It records every externally visible value together with the tolerance metadata of Section~\ref{sec:numerics}. A scoreboard sweeps device
memory at program boundaries and compares. Because the model replays the instruction stream, it also validates the encoding. A program the model refuses is a program the device must fault, and vice versa.

The suite is $34$ rows, summarized in Table~\ref{tab:rows}: a base suite of
$24$ rows and ten application benchmark rows. Every base row runs at both
synthesized configurations. This is a hard rule. A configuration that will be built must pass on its own terms.

\begin{table}[t]
\centering\footnotesize
\caption{Test inventory. The base rows pass at both configurations; the
application rows are measured at $16{\times}16$.}
\label{tab:rows}
\begin{tabular}{lll}
\toprule
class & rows & what it establishes \\
\midrule
foundation & 4 & CSR identity, model/device agreement \\
contraction & 1 & randomized shapes, modes, accumulation \\
vector & 3 & all ten operations, formats, segments \\
memory & 1 & descriptor walks, swizzles, bus errors \\
control & 2 & nested loops, co-issue, capture order \\
faults & 1 & every error code from every capable slot \\
attention & 4 & streaming, bias, ragged, bandwidth cap \\
acceptance & 1 & the frozen semantic gate \\
whole network & 1 & canonical program, both layers \\
performance & 1 & occupancy and overhead obligation \\
coverage closure & 5 & authored residue closure \\
applications & 10 & published shapes, counters read back \\
\bottomrule
\end{tabular}
\end{table}

\subsection{The semantic acceptance gate}
Three criteria were frozen before implementation:
\begin{itemize}
\item[(a)] every swept element within its composed band, zero failures;
\item[(b)] end-to-end trajectory error within the composed per-step budget;
\item[(c)] the device must \emph{denoise}, final-latent mean squared error
against a known clean latent at most one tenth that of the starting noise.
\end{itemize}

Criterion (c) is what separates a working generative accelerator from an
arithmetically defensible one. An all-zero network satisfies (a) and (b)
perfectly, since it is computed correctly and its trajectory error against
a correspondingly zero reference is nil, and it fails (c), because a network
of zeros does not denoise.

\subsection{Results}
All $24$ base rows pass at both configurations with zero errors and zero
fatals, and all ten application rows pass at $16{\times}16$; the normalization
and whole-network rows were additionally re-proved at $8{\times}8$ after the
last reference-model change.
The acceptance gate is met at both (Table~\ref{tab:gates}); the device
denoises roughly $23\times$ better than required.

\begin{table}[t]
\centering\footnotesize
\caption{Frozen semantic acceptance gate, both configurations.}
\label{tab:gates}
\begin{tabular}{lrr}
\toprule
& $16{\times}16$ & $8{\times}8$ \\
\midrule
(a) element failures & 0 / 237{,}214{,}881 & 0 / 234{,}299{,}122 \\
(b) trajectory MSE & $9.64\times10^{-6}$ & $9.79\times10^{-6}$ \\
(b) headroom to bound & $13{,}836\times$ & $13{,}278\times$ \\
(c) denoising ratio & 0.0433 & 0.0434 \\
(c) requirement & \multicolumn{2}{c}{$\leq 0.1$ (frozen)} \\
\bottomrule
\end{tabular}
\end{table}

The acceptance row was re-run after every subsequent change to any file on
its compile path. Across seven independent runs it produced bit-identical
numbers each time. We adopted that as a standing rule: a sound argument that
a change cannot affect the acceptance path is not a substitute for running
it.

\subsection{Coverage}
\begin{figure}[t]\centering
\includegraphics[width=\columnwidth]{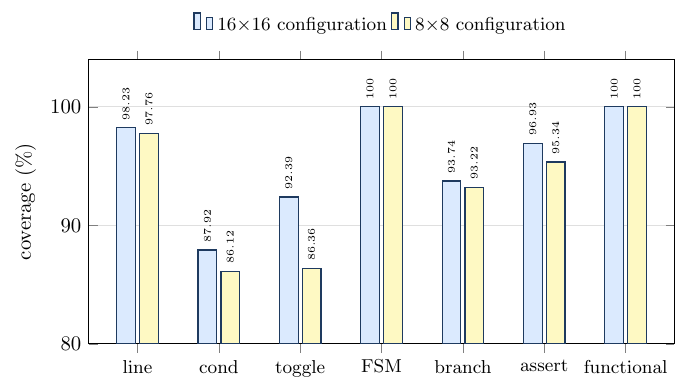}
\caption{Coverage with authored exclusions at both synthesized
configurations. Functional and FSM coverage reach $100\%$; code metrics are
reported as measured.}
\label{fig:cov}\end{figure}

Functional and finite-state-machine coverage reach $100\%$ at both
configurations (Table~\ref{tab:cov}). Code coverage is reported as measured.
The merge predates the ten application rows. Exclusions are authored by hand. Each carries a reason string naming the parameter, specification clause, error code or structural bound that makes the object unreachable. Excluding an object because it is merely hard to reach is not permitted. The exclusion files are checked under a
strict mode that fails if any excluded object is in fact covered, so a stale
exclusion cannot quietly inflate the score.

\begin{table}[t]
\centering\footnotesize
\caption{Coverage with authored exclusions; zero over-exclusion rejections.}
\label{tab:cov}
\begin{tabular}{lrr}
\toprule
metric & $16{\times}16$ & $8{\times}8$ \\
\midrule
line & 98.23 & 97.76 \\
condition & 87.92 & 86.12 \\
toggle & 92.39 & 86.36 \\
FSM & \textbf{100.00} & \textbf{100.00} \\
branch & 93.74 & 93.22 \\
assertion & 96.93 & 95.34 \\
functional & \textbf{100.00} & \textbf{100.00} \\
\midrule
overall & 95.60 & 94.11 \\
\bottomrule
\end{tabular}
\end{table}

Code coverage does not reach $100\%$ and we decline to manufacture
exclusions to make it do so. The residue is concentrated in two places. The first
is wide internal datapath buses. The second is reciprocal seed tables whose
unreached entries lie outside the reachable input domain. A directed test
could in principle reach these objects. Saying so is more useful than waiving
them.

\section{Performance}
\label{sec:perf}

\subsection{What the control path costs}
The sequencer walks a fixed stage sequence per instruction word, measured at
exactly nine cycles by a single-word zero-overhead loop of a fence
instruction. Nine cycles per word is a peak \emph{issue rate}. It is entirely hidden whenever a word's engine work exceeds nine cycles, and the minimal array injection window alone is $44$. On the
canonical program the mean engine window is $152$ cycles per word, so the
walk consumes $5.25\%$ of wall time in aggregate but only $0.006\%$ is ever
\emph{exposed} as idle. A long-standing question, whether to pipeline the
sequencer, is thereby settled by measurement: full pipelining would recover
$185$ cycles out of $3.21$ million.

\subsection{Software pipelining}
\begin{figure}[t]\centering
\includegraphics[width=\columnwidth]{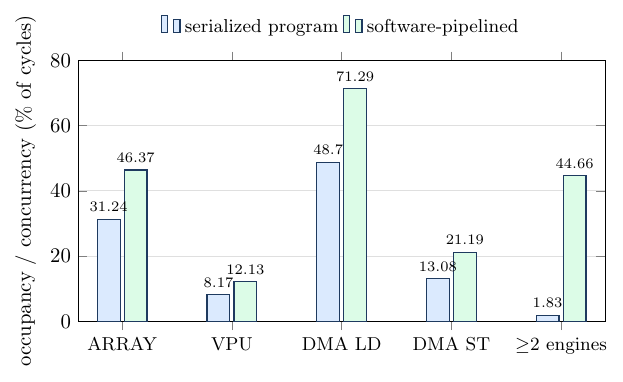}
\caption{Engine occupancy and multi-engine concurrency before and after
software pipelining the canonical program. No RTL changed; array and vector
busy cycles are bit-identical across the two runs.}
\label{fig:occ}\end{figure}

The canonical program initially fenced every loop iteration on the previous iteration's array operation fully retiring. The hardware carries an eight-deep queue of unretired operations built precisely to avoid that.
Only $1.83\%$ of cycles had two or more engines in flight. We restructured the program into a one-deep software pipeline. It prefetches the next iteration's stationary and activation tiles behind the current drain. Every removed fence was justified against a named specification rule. The result follows:

\begin{table}[t]
\centering\footnotesize
\caption{Software pipelining, canonical block program, no RTL change.}
\label{tab:softpipe}
\begin{tabular}{lrr}
\toprule
& serialized & pipelined \\
\midrule
cycles per layer & 1{,}605{,}202 & \textbf{1{,}081{,}582} \\
speedup & $1.00\times$ & $\mathbf{1.484\times}$ \\
$\geq 2$ engines in flight & 1.83\% & \textbf{44.66\%} \\
array occupancy & 31.24\% & 46.37\% \\
DMA load occupancy & 48.70\% & 71.29\% \\
non-overlapped overhead & 0.63\% & 0.26\% \\
\bottomrule
\end{tabular}
\end{table}

Two checks make the number trustworthy. Array and vector busy cycles are \emph{bit-identical} before and after, so the arithmetic did not change. The work deleted by the restructuring is bounded at ${\sim}52.5$k cycles against a $1{,}047{,}239$-cycle saving. At least $95\%$ of the gain is therefore genuine overlap. The remaining headroom is
$1.403\times$ and its limiter is named: the store's unavoidable fence
combined with engine-granular wait masks.

Attention was deliberately not pipelined. Its serialization is capacity-forced. The resident key and value groups already occupy $12$ of $16$\,KB of the stationary region, so there is no second buffer to pipeline into.

\subsection{Application benchmarks}
\label{sec:appbench}
The workload analysis identifies application classes by the hardware feature
each one requires: video~\cite{wan2025}, robot policy~\cite{chi2023diffusionpolicy}, protein structure~\cite{abramson2024af3}, 3D assets~\cite{zhang2023vecset}, guidance~\cite{ho2022cfg,esser2024sd3} and image generation~\cite{peebles2023dit}. We built one benchmark per class. Each runs a real
program on the device, reads the hardware performance counters, and reports
measured occupancy beside an analytical projection to the named production
SKU. The two are labelled separately and a projection is never presented as a
measurement. The testbench memory is small, so every benchmark runs at a
scaled shape. Table~\ref{tab:appbench} gives the measured rows.

\begin{table}[t]
\centering\scriptsize
\setlength{\tabcolsep}{3pt}
\caption{Application benchmarks, measured on the $16{\times}16$ device.
Occupancy is the fraction of cycles the engine is busy.}
\label{tab:appbench}
\begin{tabular}{llrrr}
\toprule
class & feature & cycles & array & ld.\ DMA \\
\midrule
video block (Wan 2.1) & full block, $N{=}192$ & 3{,}384{,}210 & 16.1\% & 70.3\% \\
robotics policy, $B{=}1$ & block, $N{=}64$ & 1{,}063{,}925 & 16.7\% & 72.8\% \\
protein head (AF3) & pair-bias tile & 6{,}017 & 19.7\% & 43.7\% \\
3D vecset asset & cross-attn, resident K/V & 3{,}856 & 12.1\% & 59.0\% \\
video attn, steady & skewed pipe, $G{=}8$ & 3{,}415 & 47.1\% & 45.3\% \\
guidance (SD3) & \code{V\_CFG}, $g{=}7.5$ & 51{,}647 & 0.0\% & 63.4\% \\
image (DiT-XL/2) & guidance pair, $2{\times}256$ & 9{,}365{,}022 & 15.9\% & 69.4\% \\
image, 512 px & full block, $N{=}1024$ & 30{,}259{,}588 & 15.4\% & 64.8\% \\
image attn, $d_h{=}64$ & single group & 1{,}658 & 17.6\% & 48.7\% \\
image attn, $d_h{=}72$ & single group, ragged & 2{,}620 & 15.0\% & 63.3\% \\
DDPM step (DiT) & \code{V\_AXPY} + $\sigma z$ tile & 35{,}475 & 0.0\% & 68.0\% \\
\bottomrule
\end{tabular}
\end{table}

Four results follow from the table. First, the full block is weight-bandwidth-bound at testbench shapes. Load DMA holds about $70\%$ while the array holds about $16\%$. A token sweep from $128$ to $1024$ leaves both figures unchanged. A batched guidance pair (two $256$-token images, attention per image)
scales cycles and load bytes by $1.98\times$ and leaves the array at $15.9\%$.
The cause is the emitted loop order. The GEMM sweep is stripe-outer, and every
$16$-row stripe streams each output tile's weight image into region W again,
because that region cannot hold one weight matrix. The intensity is therefore
$16$ multiply-accumulates per weight byte at every token count and batch. The
robotics row measures the same limit at its harshest point, batch-one control.

Second, the isolated attention term reaches $47\%$ array occupancy under the
skewed software pipeline, with the softmax interlock never stalling. That
interlock is a proof obligation of the design and the counter reads zero.

Third, the pair-bias and cross-attention paths run without a single stall.
Their lower array occupancy is a consequence of a sixteen-row query stripe
against a small context, which is DMA-dominated by construction. The protein
class multiplies that single pass by $24$ layers, $200$ steps and $5$ samples,
which is $24{,}000$ attention passes per query stripe per structure.

The guidance-combine row is vector work with no array involvement, and its
vector-unit occupancy of $30.2\%$ is the relevant figure. On a host this combine
is a device round trip on every sampler step; here it is one native operation.

Fourth, the image rows take their shapes from the original DiT paper~\cite{peebles2023dit}: $256$ tokens at $256$ px, $1024$ at $512$ px, head dimension $72$, and a DDPM step with learned covariance~\cite{nichol2021improved}. The
paper's head dimension of $72$ is legal through the ragged closure, and it
costs $1.58\times$ the cycles of $64$ for $1.125\times$ the useful work,
because the pad chunk is streamed and multiplied like data. The larger
finding is residency. The fused pipeline holds two K and two V groups in
region W, which is exactly full at a head dimension of $64$ and would need
$26.6$~KB at $72$. This device therefore serves head dimensions $64$ and
$128$, which cover the production video and image models, and excludes
DiT-XL/2 and PixArt~\cite{chen2023pixart}. The DDPM step with learned covariance is one vector
operation with a host-formed $\sigma z$ tile, checked against the reference.

\section{Physical Implementation}
\label{sec:results}

\begin{figure}[t]\centering
\includegraphics[width=0.92\columnwidth]{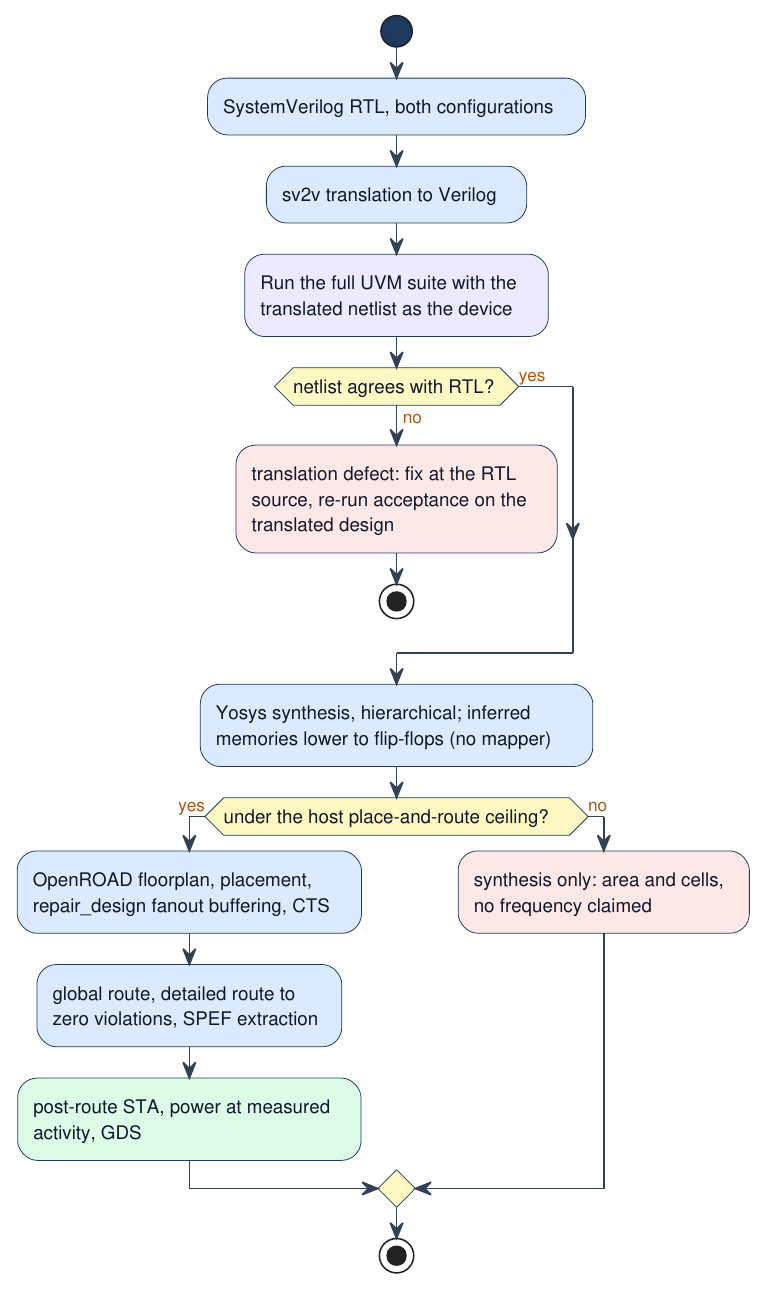}
\caption{Physical flow. The translated netlist is first fed back into the
UVM suite as the device, which validates the translation before any synthesis. Blocks under the host's place-and-route ceiling proceed to routed
layout, parasitic extraction and measured-activity power; the rest are
reported for area and cell count only.}
\label{fig:physflow}\end{figure}

WMHA was taken through the open-source SiliconCompiler/OpenROAD sky130
flow~\cite{siliconcompiler,ajayi2019openroad,wolf2013yosys,skywater2020pdk}
(Fig.~\ref{fig:physflow}). Engines were hardened standalone to routed layout with parasitic extraction;
the full chip was synthesized hierarchically. All numbers below are measured
(M) on sky130; we quote no advanced-node extrapolations.

\begin{table}[t]
\centering\scriptsize
\setlength{\tabcolsep}{3pt}
\caption{Per-engine post-route results, sky130, typical corner, $10.0$\,ns
SDC clock. Power at measured switching activity.}
\label{tab:ppa}
\begin{tabular}{lrrrr}
\toprule
engine & cells & area (mm$^2$) & fmax (M) & power \\
\midrule
\code{dix\_csr}       & 3{,}293  & 0.032 & 216.0 MHz & 4.24 mW \\
\code{dix\_pe\_cell}  & 5{,}479  & 0.050 & 153.4 MHz & 8.03 mW \\
\code{dix\_dma\_st}   & 41{,}762 & 0.273 & 128.5 MHz & 20.6 mW \\
\code{dix\_dma\_ld}   & 28{,}245 & 0.193 & 117.5 MHz & 19.0 mW \\
\code{dix\_epilogue}  & 195{,}061& 1.594 & 76.7 MHz  & 172 mW \\
\midrule
full chip (synth) & 5{,}771{,}328 & 68.355 & n/a & n/a \\
\bottomrule
\end{tabular}
\end{table}

\subsection{Host limits and what was routed}

All physical work ran on one workstation with $22$ cores and $30$\,GB of
memory. On that machine the flow has a demonstrated place-and-route ceiling
near $536$\,k instances, established on an earlier chip whose global router
finished at the edge of memory. Table~\ref{tab:pnr} lists every engine
against that ceiling. The five engines under it were routed; the four
large engines and the sequencer were not. The cell counts are dominated by
storage: the flow infers $318$ memories of two read ports and one or two
write ports, and lowers every one of them to flip-flops because the sky130
package attaches no memory mapper, so the crossbar alone becomes $530$\,k
flops.

\begin{table}[t]
\centering\scriptsize
\setlength{\tabcolsep}{4pt}
\caption{Place-and-route status of every engine against the host ceiling.}
\label{tab:pnr}
\begin{tabular}{lrl}
\toprule
engine & cells & status \\
\midrule
\code{dix\_csr}, \code{dix\_pe\_cell} & 3.3\,k, 5.5\,k & routed, GDS \\
\code{dix\_dma\_ld}, \code{dix\_dma\_st} & 28\,k, 42\,k & routed, GDS \\
\code{dix\_epilogue} & 195\,k & routed, GDS ($3$\,h $15$\,m) \\
\code{dix\_seq} & 207\,k & not converged (see text) \\
\code{dix\_axi\_arb} & 0.7\,k & pad-limited standalone \\
\code{dix\_smx} & 888\,k & exceeds host \\
\code{dix\_vpu} & 1.11\,M & exceeds host \\
\code{dix\_xbar} & 1.42\,M & exceeds host \\
\code{dix\_array} & 1.88\,M & exceeds host \\
full chip & 5.77\,M & synthesized only \\
\bottomrule
\end{tabular}
\end{table}

\subsection{Pre-layout timing is not timing}
\begin{figure}[t]\centering
\includegraphics[width=\columnwidth]{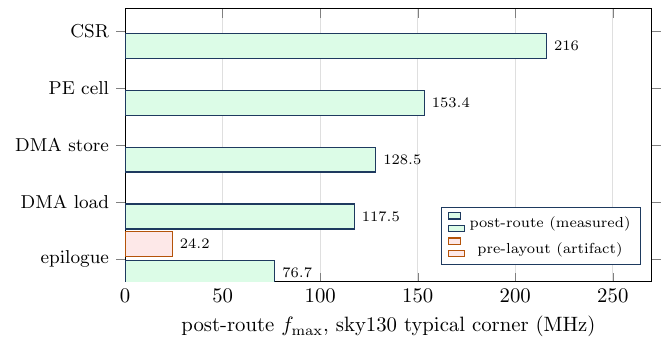}
\caption{Post-route frequency per engine, sky130 typical corner. The epilogue
bar shows the pre-layout figure for comparison: absent fanout buffering makes
pre-layout timing optimistic by $3.2\times$ on this block.}
\label{fig:fmax}\end{figure}

The most important physical-design result in this project is negative. Yosys
performs no high-fanout buffering, so a pre-layout static timing analysis on
the mapped netlist charges most of a clock period to a single minimum-drive
gate driving hundreds of loads. On the vector unit, $89.43$ of a $90.95$\,ns
path, $98\%$, sat in one \code{nor2b\_1} driving $6.98$\,pF. We do not quote pre-layout frequency figures for the four blocks affected.

The epilogue quantifies the effect directly: $24.2$\,MHz pre-layout versus
$\mathbf{76.7}$\,MHz post-route, a factor of $3.2$. Placement-time
\code{repair\_design} inserts the buffering the synthesis step never
considered. We therefore report post-route numbers only, and we recommend
that any work quoting frequencies from this class of open flow state
explicitly whether buffering has been performed.

\subsection{Measured versus assumed activity}
Power was initially computed at an assumed $0.05$ switching activity. We
replaced this with activity measured from a representative workload trace.
The result is a useful negative: the assumption was within $-8\%$ to $+4\%$
on three of four engines. The reason is structural and visible in the power breakdown. These blocks are $70$--$80\%$ internal power, of which the clock group alone is $29$--$38\%$. Only the combinational and switching terms scale with data activity. We report the measured figures. The assumed-activity shortcut happened to be defensible \emph{here}, for a reason that will not generalize to a design with a smaller clock fraction.

\subsection{Where the area goes}
The full chip synthesizes to $5{,}771{,}328$ cells and $68.355$\,mm$^2$ of
cell area in sky130, hierarchically, in $46$ minutes at a $9.7$\,GB peak. The distribution is dominated by storage and interconnect. The crossbar and banked tile memory account for roughly $25$\,mm$^2$ and the contraction array for $20$\,mm$^2$. The vector and softmax units take $8$--$9$\,mm$^2$ each. A single processing element is $0.05$\,mm$^2$. In an open-cell library without dense memory macros, a banked on-chip memory is synthesized from standard cells. A design whose interesting content is arithmetic therefore reports an area dominated by the storage around it. We report the number as measured and
note that it is not a meaningful estimate of the same architecture in a
process with compiled memories.

\subsection{Sequencer frequency is a flow artifact}
The sequencer reports $18.1$\,MHz pre-layout, which is the lowest figure in
the design and would, taken at face value, be its critical path. It is a flow artifact. The path is a ripple-carry chain $255$ logic levels deep containing $76$ majority gates. The open flow maps wide adders as ripple chains and never selects a carry-select or carry-lookahead structure. We report it as a flow floor and explicitly do not
claim it as a property of the micro-architecture. The block was also given
a $30$-hour place-and-route budget. At $40\%$ placement density detailed
routing still carried $636$\,k violations after eight iterations and was
not converging. At $18\%$ density the larger die exhausted the host's
memory during global routing. The same ripple chains that set the flow
floor make the block congestion-bound as written, so the sequencer has no
routable layout on this host in its current form.

\subsection{What we do not claim}
We report sky130 measurements only. We do not extrapolate to an advanced
node and we do not quote a TOPS figure. We also do not report a full-chip
frequency. The full chip was synthesized and not routed: at $5.77$\,M
instances it exceeds what this host can place and route. A pre-layout
full-chip number would inherit the fanout artifact documented above. The honest statement is that five engines are measured
post-route and the full chip is measured for area and cell count only.

\section{Related Work}
\label{sec:related}

Systolic and weight-stationary dataflows for dense contraction are long established~\cite{jouppi2017tpu}, and we adopt them, as is the VLIW control model~\cite{fisher1983vliw}. Our contribution is the static VLIW control path, justified by the absence of data-dependent control in diffusion inference, and the semantic acceptance criterion. Online-softmax attention descends from the streaming-softmax and memory-efficient attention literature~\cite{milakov2018online,rabe2021selfattention,dao2022flashattention}. Our variant maintains a \emph{group-wide scalar} running maximum. It also exposes the producer/consumer interlock as an architecturally visible counter, so pipeline stalls are measured.

The open-source sky130 flow~\cite{ajayi2019openroad,skywater2020pdk} is widely used for academic tapeouts and its
limitations are documented. Our contribution is a quantified statement of one
of them. We measure a $3.2\times$ gap between pre-layout and post-route
frequency on one block, attributable to absent fanout buffering. We recommend
that pre-layout figures from this flow be reported as a lower bound.

\section{Future Work}
\label{sec:future}

\subsection{Full-chip layout as a scoped project}
The full chip is synthesized and not routed, and the reason is a host, not
the design. We consider closing that gap a well-defined project on its own,
and we state here what it needs, derived from the runs that did complete.

Three pieces of design work come first. The $318$ inferred memories must be
mapped to macros or to banked single-port instances so they stop lowering
to flip-flops; that alone removes most of the $5.77$\,M cells. The array
should be assembled from the processing element already hardened here as a
macro, which leaves $479$\,k cells of array-level logic, under the ceiling
of even the present host. The sequencer's $255$-level ripple chains must be
restructured, since they are what made it congestion-bound at every density
tried. With those in place the top level becomes eleven macros and glue.

Table~\ref{tab:host} gives the machine. The memory figures extrapolate from
the run that defines the present ceiling, $536$\,k instances at the edge of
$30$\,GB, which is about $56$\,GB per million final instances, with a
$1.6\times$ margin. The times extrapolate from the two measured engine runs,
$52$ minutes at $42$\,k cells and $3$\,h $15$\,m at $195$\,k, with a
$1.2$ exponent for congestion growth. Memory decides whether a run finishes
at all; cores only shorten the wait, and detailed routing scales to about
$32$ threads.

\begin{table}[t]
\centering\scriptsize
\setlength{\tabcolsep}{4pt}
\caption{Estimated host requirement per block, standalone, extrapolated from
measured runs on the $30$\,GB host.}
\label{tab:host}
\begin{tabular}{lrrr}
\toprule
block & memory at edge & recommended & wall time \\
\midrule
\code{dix\_smx} & 50\,GB & 96\,GB & ${\sim}20$\,h \\
\code{dix\_vpu} & 62\,GB & 128\,GB & ${\sim}26$\,h \\
\code{dix\_xbar} & 79\,GB & 128\,GB & ${\sim}35$\,h \\
\code{dix\_array}, flat & 105\,GB & 192\,GB & ${\sim}49$\,h \\
\code{dix\_array}, PE as macro & 27\,GB & 64\,GB & ${\sim}10$\,h \\
\code{dix\_seq}, $18\%$ density & 35\,GB & 64\,GB & ${\sim}13$\,h \\
full chip, flat, as is & 323\,GB & 768\,GB & ${\sim}8$\,days \\
\bottomrule
\end{tabular}
\end{table}

A machine with $256$\,GB of memory, $32$ to $64$ cores and $1$ to $2$\,TB
of local solid-state storage carries the whole macro-hierarchical path:
every engine fits with margin, two engines can run concurrently, and the
sum of the per-block times is about one week sequentially or three to four
days with two engines in flight. The flat full chip as it stands would need
roughly $768$\,GB and eight days, and we expect it to behave like the
sequencer and fail to converge, so we do not recommend that path even on
adequate hardware.

\subsection{Specification revisions the benchmarks motivate}
Two application findings are sizing decisions for the next revision of the
frozen specification. First, the fused attention pipeline's resident ring is
exactly region W at a head dimension of $64$ and needs $26.6$\,KB at $72$,
so the original DiT head dimension is served only by the single-group path;
doubling region W or adding a three-chunk reach are the candidate remedies.
Second, the block program is weight-bound at every token count and batch
because the GEMM sweep is stripe-outer and region W cannot hold one weight
matrix; weight residency across stripes, or taller stripes at a larger
accumulator footprint, would let batching amortise the weight stream as the
workload analysis assumed.

\subsection{Verification residue}
Code coverage stands at $95.6\%$ and $94.1\%$ with authored exclusions and
we have declined to waive the remainder. The residue is reachable by
directed stimulus and is listed. The coverage merge predates the ten application rows, and a full re-run of
the base suite at both configurations with those rows folded into the merge
is the next verification step.

\section{Conclusion}

WMHA is a latency-first diffusion-transformer inference accelerator. Its architecture follows from a property of the workload. Diffusion inference has no autoregressive recurrence, so its schedule is fully known at compile time. A static VLIW control path with explicit dependences is therefore the natural fit. The design is specified in a frozen document. It is verified against a double-precision reference at both synthesized configurations, with $100\%$ functional and finite-state-machine coverage. It is taken to routed layout with parasitic-annotated timing and measured-activity power. It meets a semantic acceptance criterion, frozen
before implementation, by a factor of $23$.

The performance result we consider most instructive required no hardware change at all. Restructuring the instruction schedule into a one-deep software pipeline raised multi-engine concurrency from $1.83\%$ to $44.66\%$ of cycles. The measured speedup is $1.484\times$, with the arithmetic proven unchanged by bit-identical busy counts. The hardware had been built to
overlap; the program was declining to.

The physical results are sky130 measurements and nothing more. Five engines
are routed with parasitic-annotated timing and measured-activity power. The
full chip is synthesized, and the host limit that stopped its layout is
quantified together with the machine and the design work that would remove
it. Closing that gap is the natural next project.

\bibliographystyle{IEEEtran}
\bibliography{refs}

\end{document}